\documentclass[preprint2]{aastex}

\slugcomment{Draft B1}

\shorttitle{M-I coupling at ultra-cool dwarfs}
\shortauthors{Nichols et al.}

\begin{document}

\title{Origin of electron cyclotron maser-induced radio emissions at \\ ultra-cool dwarfs: magnetosphere-ionosphere coupling currents}

\author{J.~D.~Nichols, M.~R.~Burleigh, S.~L.~Casewell, S.~W.~H~Cowley, and G.~A.~Wynn}
\affil{Department of Physics and Astronomy, University of Leicester, Leicester, LE1~7RH, UK}
\email{jdn@ion.le.ac.uk}

\and

\author{J.~T.~Clarke and A.~A.~West}
\affil{Center for Space Physics, Boston University, Boston, MA 02215, USA}

\begin{abstract}
A number of ultra-cool dwarfs emit circularly polarised radio waves generated by the electron cyclotron maser instability.  In the solar system such radio is emitted from regions of strong auroral magnetic field-aligned currents.  We thus apply ideas developed for Jupiter's magnetosphere, being a well-studied rotationally-dominated analogue in our solar system, to the case of fast-rotating UCDs.  We explain the properties of the radio emission from UCDs by showing that it would arise from the electric currents resulting from an angular velocity shear in the fast-rotating magnetic field and plasma, i.e. by an extremely powerful analogue of the process which causes Jupiter's auroras. Such a velocity gradient indicates that these bodies interact significantly with their space environment, resulting in intense auroral emissions. These results strongly suggest that auroras occur on bodies outside our solar system.  
\end{abstract}



\section{Introduction}

Ultra-cool dwarfs (UCDs) are objects with spectral type later than M7, comprising the lowest mass stars and brown dwarfs.  Twelve of these (out of $\sim$200 observed) have been found to be intense sources of radio emissions with spectral luminosities typically of order a $\mathrm{MW\;Hz^{-1}}$ \citep{berger:2001aa, berger:2009aa, berger:2006aa, hallinan:2006aa,hallinan:2007aa, hallinan:2008aa,antonova:2007aa,antonova:2008aa, phan-bao:2007aa,mclean:2011aa,mclean:2012aa, route:2012aa}.  The unpolarised component probably includes synchrotron emission, but the radio emission from a number of these, which have fast ($\mathbf{\sim}$2~h) rotation periods and strong ($\sim$$0.1$~T) magnetic fields, has been shown to be highly circularly polarised and modulated at the bodies' rotation periods \citep[e.g.][]{hallinan:2006aa}.  The polarised nature of the emission implicates the electron cyclotron maser instability (CMI) as the source mechanism \citep{wu79a, treumann:2006aa}. \cite{yu:2011aa} modelled the radio emission, assuming it is generated by unaccelerated precipitating hot plasma trapped on coronal loops.  However, this process would be likely limited to short-lived flare-type events, does not directly result in increased radio power with faster rotation as is observed \citep{mclean:2012aa}, and the CMI is known to be generated at much greater efficiency when electrons are accelerated down magnetic field lines by field-aligned voltages \citep{treumann:2006aa}.  The presence of sustained CMI-generated radio emissions is thus instead strongly suggestive of the existence of quasi-steady auroral magnetic field-aligned currents.  Indeed, all CMI source regions observed in situ at planets in the solar system exhibit accelerated electron populations, and radio emissions generated by the CMI have been extensively shown to be closely associated with auroral emission, caused when downward-precipitating electrons impact the atmosphere \citep{treumann:2006aa, clarke09a,lamy09a,lamy:2010aa, nichols10a,nichols10b}. \cite{yu:2012aa} have recently developed their analysis to consider the results of an accelerated population of electrons, and their simulations indicate that such populations could produce the observed UCD emissions.   \\

Field-aligned currents arise from a divergence in field-perpendicular currents, which are, through $\mathbf{E}=-\mathbf{v}\times\mathbf{B}$, driven by plasma velocities relative to the neutrals in the conducting outer layer of the atmosphere.  Thus, strong field-aligned currents flow when there is a sharp gradient in this departure from corotation, giving rise to a strong divergence in the field-perpendicular currents. We may therefore infer from the observation of CMI-generated radio emissions from UCDs that such angular velocity gradients with auroral currents exists in the magnetospheres of these objects.  At planets in the solar system angular velocity gradients occur near the boundary between open and closed field lines, as at Earth and Saturn, or are due to centrifugally-driven outflow of internally-generated plasma, as at Jupiter.  \cite{schrijver:2009aa} presented the hypothesis that the radio emission from UCDs could result from a Jupiter-like current system on the basis of a simple dimensional scaling relation, but did not consider quantitatively the magnetosphere-ionosphere (M-I) coupling current system.  Here we have developed a simple axisymmetric model of the currents arising from departure from rigid corotation in the magnetospheres of UCDs, based on that used previously to study Jupiter's auroral oval \citep{hill79,hill01, cowley01,cowley02,nichols03,nichols04, nichols05,cowley05a,nichols11b} and to estimate the radio luminosity of fast-rotating Jupiter-like exoplanets \citep{nichols11a,nichols:2012aa}.  We explain the properties of the radio emission from UCDs by showing that it would arise from the electric currents resulting from an angular velocity shear in the fast-rotating magnetic field and plasma, i.e.\ by an extremely powerful analogue of the process which causes Jupiter's or Saturn's auroras.  Such a velocity gradient indicates that these bodies interact significantly with their space environment, resulting in intense auroral emissions.\\ 

\section{Theoretical background} 
\label{sec:theory}

\subsection{Basic assumptions}
\label{sec:assumps}

 For the purposes of development of the model, we assume that a jovian picture of an ultra-cool dwarf (UCD) is appropriate, i.e.\ that the UCD's magnetic field is generated deep within its interior, such that the external field generated by the body is mainly dipolar, and that a non-conducting atmosphere surrounds the body, the upper portion of which is maintained to some degree conductive by photon and/or corpuscular impact from the outside.  The atmospheres of cool UCDs are not expected to be strongly ionised, although it has been suggested that dust clouds may induce some degree of ionisation \citep{helling:2011aa,helling:2011ab}.  At solar system planets a number of sources of conductivity are known besides solar X-ray and EUV radiation, such as galactic cosmic rays and auroral electron precipitation \citep{rycroft:2008aa}.  The latter is particularly important, since it results in a positive feedback which greatly increases the intensity of the auroral currents. At Jupiter, for example, modeling work indicates that the feedback from auroral precipitation increases the conductance from a residual value of $\sim$0.01~mho  (where 1 mho $\equiv$ 1 siemens) generated e.g.\ by solar X-ray and EUV irradiation to values of up to 10~mho \citep{strobel83,millward02}, such that the feedback plays a dominant role in influencing the location and intensity of the M-I  coupling currents \citep{nichols04}. Thus we suggest that given some small initial conductivity generated by a source mentioned above, the auroral current system would amplify itself via the feedback from the auroral precipitation.  The exact details of such feedback requires ionospheric modeling that is beyond the scope of the present paper, but suffice it to say that due to this effect the low temperatures of UCDs do not necessarily prohibit the flow of strong M-I coupling currents.  In the absence of such an ionospheric model for UCDs, in the present work we simply take the Pedersen conductance $\Sigma_P$ to be a constant, discussed further below.\\

A principal assumption is that the spin and magnetic axes are roughly co-aligned, since this is the configuration that has been most studied for bodies in our solar system, although we note that it is probable that only $\sim$50\% of UCDs exhibit this polarity.  With reversed field polarity, the associated M-I coupling current system described below would also be reversed, such that the upward current, which in Fig. 1 is strongly peaked at 15$^\circ{}$, would instead be widely distributed at high latitudes, and the strong downward currents would be easily carried by upward-going ionospheric electrons, leading to little auroral and radio emissions.  We also assume approximate axisymmetry about the magnetic axis, such that the flows in the ionosphere are considered to be wholly azimuthal about that axis, and can thus be described in terms of departure from corotation with the neutral atmosphere as a function of colatitude $\theta_i$ from the magnetic axis. Such a simplification seems appropriate for representing to lowest order the dynamics of a fast rotating magnetosphere, although we note that some processes related to an open magnetosphere will consequently be excluded from the model, such as `cusp-like' or `substorm-like' processes which would be ordered in azimuth with respect to the object's velocity through the surrounding medium.  We further assume for simplicity that departures of the upper neutral atmosphere from rigid corotation due to ion-neutral drag also take the form of winds that are approximately axisymmetric about the magnetic axis, thus neglecting the effect of the Coriolis force on such motions in cases where the spin and magnetic axes are significantly inclined.\\

It has been suggested by \cite{kuznetsov:2012aa} that the short duration of the spikes in the CMI-induced radio emissions is indicative of an active sector in the magnetosphere of a UCD.  We suggest, however, that the observed short duty cycles of $\leq$0.15 for UCD radio emissions \citep{hallinan:2006aa} are not inconsistent with an essentially axisymmetric field-aligned current distribution.  Jupiter's main auroral oval is to first order described by an axisymmetric annulus of isotropic emission $\sim$$1^\circ{}$ wide located $\sim$$15^\circ{}$ from the magnetic pole, which is tilted by $\sim$10$^\circ{}$ from the spin axis \citep{grodent03b, clarke04, nichols09b}.  Therefore, as the dipole axis cones around the spin axis, from a viewpoint near the planet's spin equatorial plane over half of the auroral oval is visible over roughly a third of the planet's rotation period \citep{nichols07}.  The associated radio emission, however, is strongly beamed, into $\sim1.6$~sr in the case of the HOM and DAM emissions \citep{zarka04a}.  Hence, the radio emission peaks strongly as the beam rotates into view, and the FWHM of the occurrence probability distribution of the `Source A' DAM emission (i.e.\  that used to determine Jupiter's System-III rotation period of $\sim$9~h 55~min) is $\sim$$50^\circ{}$, i.e.\ an apparent duty cycle of $\sim$$0.14$ \citep{higgins:1996aa}.  This is consistent with the duty cycles of $\leq0.15$ for UCDs.  Having said this, it is important to recognise that axisymmetry is simply a lowest order approximation. Saturn's radio emissions, for example, are observed from all local times but are brightest in the morning sector \citep{lamy09a}, despite the internal planetary magnetic field being axisymmetric to within detection limits \citep{burton:2010aa}. Finally, as has also been suggested by \cite{berger:2009aa}, it is probable that as is observed at Jupiter the magnetic poles may not lie on respective antipodes, due to the non-dipolar terms in the internal field, such that the radio emission from the two hemispheres would not necessarily be expected to be observed 180$^\circ$ apart in phase. \\

\subsection{Current system equations}
\label{sec:currents}

We consider three angular velocities in the computation of the M-I coupling currents.  The first is the angular velocity of the deep interior of the UCD $\Omega_{UCD}$, taken initially to be $8.7\times10^{-4}\mathrm{\;rad\;s^{-1}}$ in conformity with the observed periods of $\sim$2~h for radio-bright UCDs.  The second is the angular velocity of the plasma above the ionosphere $\omega$, taken to be constant on each flux tube, and the third is the angular velocity of the atmospheric neutrals in the Pedersen layer of the ionosphere $\Omega_{UCD}^*$, which is expected to lie somewhere between $\Omega_{UCD}$ and $\omega$ due to the frictional drag of ion-neutral collisions \citep{huang89, millward05}.  In such a case we have

\begin{equation}
	(\Omega_{UCD}-\Omega_{UCD}^*)=k(\Omega_{UCD}-\omega)\;\;,
	\label{eq:slip}
\end{equation}

\noindent where $0<k<1$.  The actual value of $k$ is somewhat uncertain for Jupiter, with modelling work indicating a value of $\sim$0.5  \citep{millward05}, which is employed in the previous jovian modelling work discussed and also now adopted here, although the results are not expected to be sensitively dependent on reasonable choices.  We should note that in the following work we consider only sub-rotation of the plasma with respect to the UCD.  For Jupiter, super-rotation of the magnetospheric plasma may occur following strong solar wind-induced compressions of the magnetosphere, with a corresponding reversal of the current system described below \citep{hanlon04b,cowley07, nichols07}, but in the steady state the plasma sub-rotates, associated with the transfer of angular momentum from the atmosphere to the external medium. \\

The equatorward-directed height integrated ionospheric Pedersen current is then given by

\begin{equation}
	i_P = \Sigma_P E_i=\Sigma_P\rho_i(\Omega_{UCD}^* - \omega)B_i\;\;,
	\label{eq:ip1}
\end{equation}

\noindent where $E_i$ is the electric field in the rest frame of the neutral atmosphere, $\rho_i = R_{UCD} \sin\theta_i$ is the distance from the spin axis  (where we assume initially the UCD has a radius $R_{UCD}$ equal to Jupiter's equatorial 1-bar level radius $R_J = 71,373$~km since all bodies of roughly solar composition from giant planets through to the very lowest mass stars have radii roughly similar to Jupiter).  Employing equation~\ref{eq:slip} in equation~\ref{eq:ip1} and introducing the `effective' Pedersen conductance $\Sigma_P^*$, reduced from the true value by a factor of $(1-k)$ owing to the slippage of the neutral atmosphere from rigid corotation, yields

\begin{equation}
	i_P = \Sigma_P^*\rho_i(\Omega_{UCD}-\omega)B_i\;\;.
	\label{eq:ip2}
\end{equation}

\noindent As mentioned above, we simply take a constant value for $\Sigma_P^*$, initially 0.5~mho, which is similar to the values employed in previous Jupiter modelling works, and, as shown below, yields radio power values in agreement with observations of UCDs.  However, in the results below we also examine the effect of different values of this parameter.  The total Pedersen current at a given co-latitude is then found by integrating in azimuth, such that

\begin{equation}
	I_P = 2\pi\rho_ii_P = 2\pi \Sigma_P^* \rho_i^2(\Omega_{UCD}-\omega)B_i\;\;,
	\label{eq:totip}
\end{equation}

\noindent the divergence of which gives the field-aligned current density just above the ionosphere, i.e. for the northern hemisphere\ 

\begin{equation}
	j_{\|i} = -\frac{1}{2\pi \rho_i^2 \sin \theta_i}\frac{dI_p}{d\theta_i}\;\;.
	\label{eq:jpari}
\end{equation}

\subsection{Field-aligned acceleration and energy flux}
\label{sec:fav}

The upward-directed field-aligned current computed in the previous section can not, in general, be carried simply by precipitating magnetospheric electrons alone, and must be driven by downward acceleration of those electrons by a field-aligned voltage.  The maximum field-aligned current density that can be carried by an unaccelerated isotropic Maxwellian population is

\begin{equation}
	j_{\|i0} = en\left(\frac{W_{th}}{2\pi m_e}\right)^{1/2}\;\;,
	\label{eq:jpari0}
\end{equation}

\noindent where $e$, $m_e$, $n$ and $W_{th}$ are the charge, mass, number density and thermal energy of the electron source population, respectively, the latter being equal to equal to $k_BT$, where $k_B$ is Boltzmann's constant and $T$ is the temperature.  In common with previous works we use the values for the high latitude hot plasma at Jupiter, i.e.\ $n=0.01~\mathrm{cm}^{-3}$ and $W_{th}=2.5$~keV \citep{scudder81}, which, as shown below yield results consistent with observations, but we also examine the effects of wide ranges of values for these parameters.  The corresponding unaccelerated kinetic energy flux is

\begin{equation}
	E_{f0} = 2enW_{th}\left(\frac{W_{th}}{2\pi m_e}\right)^{1/2}\;\;.
	\label{eq:ef0}
\end{equation}

\noindent For field-aligned current densities larger than $j_{\|i0}$, a field-aligned voltage is required.  For Earth, the current-voltage relation derived using the kinetic theory of \cite{knight73} is applicable, but for more powerful systems in which either the source electron population is very energetic initially, or becomes so following acceleration by voltages comparable to or exceeding the electron rest mass energy ($\sim$511~keV), Cowley's (2006) \nocite{cowley06b} relativistic extension to Knight's theory becomes appropriate.  In the results shown in this paper it is clear that the relativistic theory is required in the case of M-I coupling current systems at UCDs, such that we employ the relativistic current-voltage relation given by

\begin{equation}
	\left(\frac{\ensuremath{j_{\|i}}}{\ensuremath{j_{\|i\circ}}}\right)=
	1 + \left(\frac{e\Phi}{W_{th}}\right)+
	\frac{\left(\frac{e\Phi}{W_{th}}\right)^2}{2\left[\left(\frac{m_ec^2}{W_{th}}\right)+1\right]}\;\;,
	\label{eq:phi}
\end{equation}

\noindent where $c$ is the speed of light and $\Phi$ is the minimum voltage required to drive the current $j_{\|i}$ at ionospheric altitude, applicable in the limit that the accelerator is located at high altitude along the field line (where $B \ll B_i$).  The corresponding precipitating electron energy flux is given by

\begin{equation}
		\left(\frac{E_f}{E_{f0}}\right) = 
		1 + \left(\frac{e\Phi}{W_{th}}\right) + 
			\frac{1}{2}\left(\frac{e\Phi}{W_{th}}\right)^2 +	
			\frac{\left(\frac{e\Phi}{W_{th}}\right)^3}
			{2\left[2\left(\frac{m_ec^2}{W_{th}}\right)+3\right]}\;\;.
	\label{eq:ef}
\end{equation}

\noindent Equations~\ref{eq:jpari0}-\ref{eq:ef}, along with values for the electron source population number density and temperature as discussed further below, are thus used to determine the precipitating energy flux resulting from the current system discussed above.  The total precipitating power for each hemisphere $P_e$ is then obtained by integration of \ensuremath{E_f} over the hemisphere, i.e.\ 

\begin{equation}
	P_e=\int_0^{90}2\pi R_{UCD}^2\sin\theta_i \;E_f \;d\,\theta_i\;\;,
	\label{eq:pe}
\end{equation}

\noindent and we finally obtain the spectral luminosity using

\begin{equation}
	L_r=\frac{P_e}{100 \: \Delta \nu}\;\;.
	\label{eq:pr}
\end{equation}

\noindent where we assume that the beam from only one hemisphere is observable at any one time and we take the electron cyclotron maser instability to have a generation efficiency of \ensuremath{\sim}1\%, a value observed at both Jupiter and Saturn. Specifically, at Saturn this value was directly measured during the traversal of the Cassini spacecraft through the source region of the Saturn Kilometric Radiation (SKR) \citep{lamy11a}, and at Jupiter the radio power output is $\sim$1\% of the total precipitating electron power, as measured from observations of the UV aurora and computed theoretically \citep{gustin04a,clarke09a,cowley05a}.  The bandwidth $\Delta \nu$ is assumed to be equal to the electron cyclotron frequency in the polar ionosphere, i.e.\ 

\begin{equation}
	\Delta\nu=\frac{eB_i}{2\pi m_e}\;\;,
	\label{eq:delnu}
\end{equation}

\noindent an approximation validated by observations of solar system planets.  For example, \cite{zarka98a} argue that the jovian b-KOM, HOM and DAM emissions are a single radio component, thus with a frequency range of $\sim$10~kHz to $\sim$40~MHz, and at Saturn the SKR extends from a few kHz to $\sim$1200~kHz.  This arises because the CMI occurs at all points below the field-aligned voltage drop, situated at least a few planetary radii up the field line, and the top of the ionosphere at 1 planetary radius.  The emission is generated at the local electron cyclotron frequency, such that the bandwidth is then determined by the difference between the field strengths at these two locations.  The strength of a dipole field drops off approximately as cube of the distance along the field line, such that at altitudes of only $\sim$1 planetary radius up the field line the frequency drops by an order of magnitude.  The bandwidth is thus reasonably well represented by the high frequency cutoff, which is also the frequency used to measure the magnetic field strength in the polar region.

\begin{figure}
\noindent\includegraphics[width=83mm]{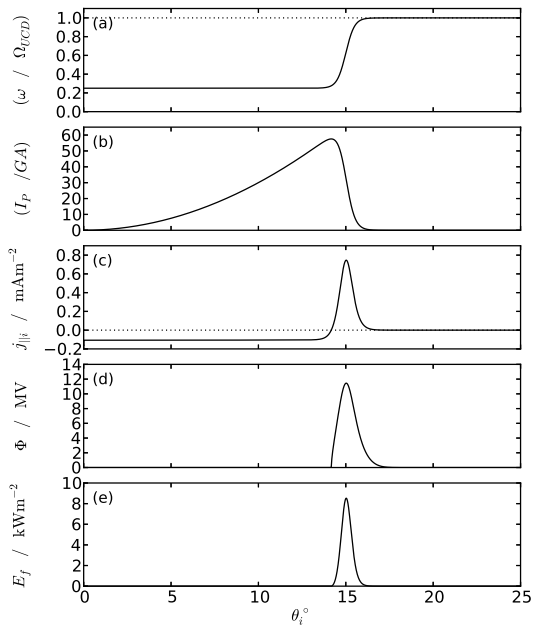}
\caption{
Figure showing profiles of model M-I coupling current system parameters.  Specifically we show (a) the plasma angular velocity normalised to that of the UCD $(\omega/\Omega_{UCD})$, (b) the azimuth-integrated ionospheric Pedersen current $I_P$ in GA, (c) the field-aligned current density at the top of the ionosphere $j_{\|i}$ in $\mathrm{mA\;m^{-2}}$, (d) the minimum field-aligned voltage required to drive the current plotted in panel (c) in MV, and (e) the resulting precipitating electron energy flux $E_f$ in $\mathrm{kW\;m^{-2}}$, all plotted versus ionospheric co-latitude $\theta_i$.  The Pedersen conductance employed is $\Sigma_P^*=0.5$~mho and the source electron number density and thermal energy are jovian values of 0.01~cm$^{-3}$ and 2.5~keV. The horizontal dotted lines in panels (a) and (c) indicate values of $(\omega/\Omega_{UCD})=1$ and $j_{\|i}=0$~$\mathrm{mA\;m^{-2}}$, respectively.
}
\label{fig:profs}
\end{figure}

\subsection{Angular velocity profile}

We now discuss the model for the plasma angular velocity used in this paper.  Following previous work \citep{cowley05a}, the gradient in the angular velocity is conveniently represented by a tanh function with respect to co-latitude, given by


\begin{eqnarray}
	\lefteqn{\left(\frac{\omega}{\Omega_{UCD}}\right)= \left(\frac{\omega}{\Omega_{UCD}}\right)_L + }\nonumber \\ &\frac{1}{2}\left(1 - \left(\frac{\omega}{\Omega_{UCD}}\right)_L\right) \left(1+\tanh\left(\frac{\theta_i - \theta_{ic}}{\Delta\theta_i}\right)\right)\;\;,
	\label{eq:omega}
\end{eqnarray}

\noindent where $(\omega/\Omega_{UCD})_L$ is the low value to which the angular velocity transitions near the magnetic pole from rigid corotation at lower latitudes, and $\theta_{ic}$ and $\Delta\theta_i$ are the latitudinal centre and half-width of the transition region, respectively.  We initially employ values for these parameters which give a Jupiter-like angular velocity profile, which on that planet generates the field-aligned currents which cause the main auroral oval and the HOM, b-KOM and non-Io-DAM radio emissions, i.e.\  $(\omega/\Omega_p)_L = 0.25$, $\theta_{ic}=15^\circ$ and $\Delta\theta_i = 0.5^\circ$. The actual form of the angular velocity gradient depends on the physical processes causing the shear, as is examined further in the Discussion, and in the results below we also examine the effects on the radio power of taking different values for these parameters. \\

\section{Results} 
\label{sec:results}
	
\subsection{Results with representative values of system parameters}
\label{sec:spot}

In this section we present results of the model using appropriate spot values for the various model parameters as discussed above, before going on in the next section to examine ranges in the parameter space.  Figure~\ref{fig:profs} shows representative results, in which we employ a uniform ionospheric field strength of 0.3~T and rotation period of 2~h, in conformity with typical values obtained from the bandwidth and modulation of the UCD radio emissions \citep{hallinan:2008aa}.  First, Figure~\ref{fig:profs}a shows the angular velocity profile employed, given by equation~\ref{eq:omega}.  Moving from large to small co-latitudes (or equivalently on field lines which thread the equatorial plane at increasing radial distances), the plasma angular velocity initially near-rigidly corotates, and transitions over $\sim$1$^\circ{}$ at $15^\circ{}$ co-latitude to a quarter of rigid corotation at smaller co-latitudes.  The resulting equatorward azimuth-integrated ionospheric field-perpendicular (Pedersen) current given by equation~\ref{eq:totip} is shown in Figure~\ref{fig:profs}b, increasing from small values near the pole to a value of $\sim$58~GA, before falling rapidly due to the plasma angular velocity gradient shown in Figure~\ref{fig:profs}a.  The resulting field-aligned current given by equation~\ref{eq:jpari} is shown in Figure~\ref{fig:profs}c, which is negative, i.e.\ downward, near the pole and switches to a significant peak of positive, i.e.\ upward, field-aligned current values, reaching amplitude $\sim$$0.7~\mathrm{mA\;m^{-2}}$, centred on $15^\circ{}$.  This is the region of auroral field-aligned current that sustains the unstable electron distributions responsible for the CMI and thus the radio emissions. The required field-aligned voltage given by equation~\ref{eq:phi} is then shown in Figure~\ref{fig:profs}d, peaking at $\sim$11~MV, and the corresponding precipitating electron energy flux given by equation~\ref{eq:ef} is shown in Figure~\ref{fig:profs}e, peaking at $\sim$$9~ \mathrm{kW\;m^{-2}}$.  Integrating this energy flux and converting to spectral luminosity as discussed above yields a radio power output  of $\sim$$1.1~\mathrm{MW\;Hz^{-1}}$, i.e.\ consistent with the typical observed radio luminosities of UCDs.  We also note that the bandwidth of the radio emission, equal to the electron cyclotron frequency of $\sim$8.4~GHz, is perforce similar to that observed, since the ionospheric field used is determined from the measured radio frequencies.\\

\begin{figure}[t]
\noindent\includegraphics[width=83mm]{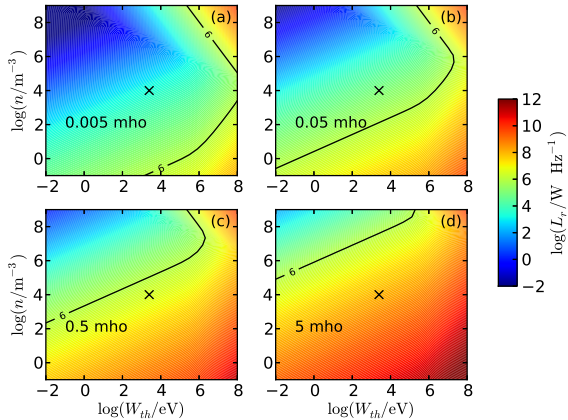}
\caption{
Plots showing the spectral luminosity computed in the model for different pairs of values of the source electron population number density $n$ and thermal energy $W_{th}$.  In panels (a-d) we use Pedersen conductances of $\Sigma_P^*=$~0.005, 0.05, 0.5, and 5 mho, respectively.  The crosses indicate the jovian values employed in Section~\ref{sec:fav}, and the black contour indicates where $L_r=1\;\mathrm{MW\;Hz^{-1}}$, comparable to observed values. 
}
\label{fig:contours}
\end{figure}
\subsection{Parameter space investigation} 
\label{sec:params}

In the above section we have used reasonable values for a number of system parameters based on observations at Jupiter in order to demonstrate the model and indicate how an angular velocity gradient leads to radio luminosity that, for the case of fast-rotating UCDs, is comparable to that observed.  However, the only parameters appropriate to this model that are known from observations, besides the radio power, are the polar magnetic field strength and rotation rate of such objects.  The other parameters, such as the form of the angular velocity profile, the density and temperature of the source electron population and the ionospheric conductance are not known.  Here we therefore examine ranges of these parameters to determine what values are consistent with the observed radio emission.  We first examine the effects of the conductance and source electron population parameters, given the angular velocity profile given by equation~\ref{eq:omega}, and we go on to investigate the effect of changing the form of the angular velocity profile. \\

In each panel of Figure~\ref{fig:contours} we show the radio spectral luminosities obtained from the model using different pairs of values for the source electron thermal energy $W_{th}$ and number density $n$.  The values adopted above obtained from Voyager observations of the plasma outside the current sheet at Jupiter, i.e.\  $~$2.5~keV and $\sim$0.01~$\mathrm{cm^{-3}}$, respectively, are shown by the crosses in each panel.  The low density arises since the cold plasma is centrifugally confined to the equatorial plane, and only the hot, rarefied plasma population has a centrifugal confinement scale height large enough to populate the entire flux tube \citep{hill76a, caudal86}.  At high latitudes, the low beta plasma, combined with the field-aligned voltage and loss of particles to the ionosphere, is then favourable for generation of the CMI.  We note that previous estimates of the electron number density and temperature on the flux tubes producing the CMI at UCDs have been obtained, i.e.\  1.25-5$\times10^5\;\mathrm{cm^{-3}}$ and 1-5$\times10^7$~K (equivalent to $\sim$0.86~keV) by \cite{yu:2011aa}, but these assume that the CMI is produced by an unaccelerated loss-cone population.  We suggest, however, that at fast rotating UCDs, at which the magnetospheric plasma is likely to be significantly more centrifugally confined than at Jupiter, the high latitude plasma parameters may possibly be closer to the jovian case.  In light of the uncertainty surrounding the source plasma population, we show the results of the model using the angular velocity profile given by equation~\ref{eq:omega}, for wide ranges of these parameters, i.e.\ 5 orders of magnitude either side of the jovian values.  In panels (a-d) we employ $\Sigma_P^*=$~0.005, 0.05, 0.5, and 5 mho, respectively, reflecting the uncertainty in the Pedersen conductance.  The contour labelled ``6'' indicates where the radio luminosity is equal to 1~$\mathrm{MW\;Hz^{-1}}$, representative of the observed luminosities.  \\

\begin{figure}
\noindent\includegraphics[width=83mm]{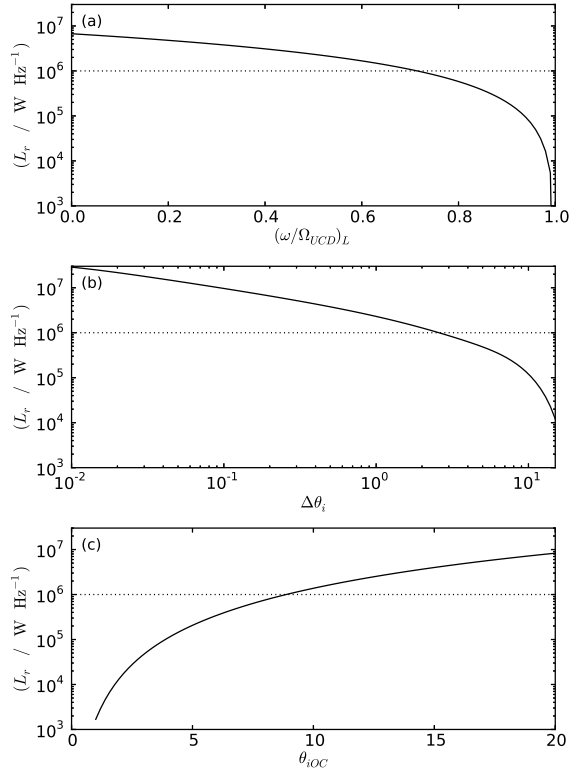}
\caption{
Plots showing the spectral luminosity computed in the model for different values of the angular velocity profile parameters.  These are (a) the low value to which the angular velocity transitions from rigid corotation $(\omega / \Omega_p)_L$, (b) the latitudinal half-width of the transition region $\Delta\theta_i$, and (c) the centre of the transition region $\theta_{ic}$. The horizontal dotted lines are located at $L_r = 1\;\mathrm{MW\;Hz^{-1}}$.  As previously, source electron population parameters are fixed at jovian values.  
}
\label{fig:params} 
\end{figure}

\begin{figure}[h]
\noindent\includegraphics[width=83mm]{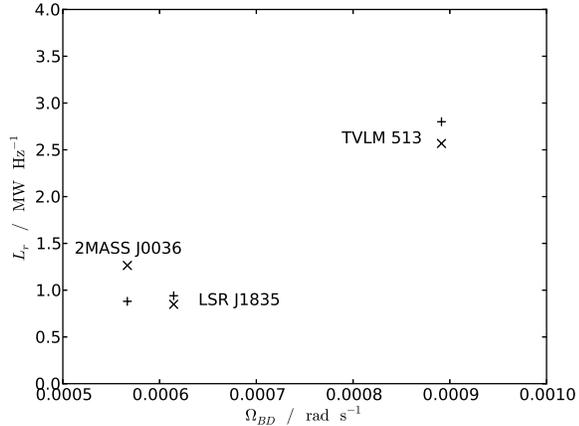}
\caption{
Figure showing modelled (pluses) and observed (crosses) radio luminosities of the three confirmed UCD sources of CMI-generated radio emissions versus each UCD's angular velocity.  Model parameters are given in Table~\ref{tab:ucds}
}
\label{fig:cases}
\end{figure}

It is apparent that over the majority of the parameter space plotted, the luminosity increases with decreasing $n$ and increasing $W_{th}$, with a change of slope as the relativistic terms dominate above $W_{th}\simeq511$~keV.  Where the slope reverses in the top right corner of each panel the field-aligned current can be carried by an unaccelerated population.  For simplicity, in this region we have assumed that in the auroral region the loss-cone is filled by strong pitch angle diffusion, and that the required net current is then created by a corresponding upward flux of electrons.  We have also taken the same CMI generation efficiency of 1\% as for the accelerated population, despite the less favourable loss-cone distribution in this case.  Both these assumptions will likely lead to overestimates of the radio flux from this region.  We also note that the previous estimates of the source population parameter values discussed above \citep{yu:2011aa} are off the top of each panel, thus clearly in the unaccelerated regime.  In any case, it is apparent in the accelerated region that a wide range of parameter values are able to produce the required radio luminosity, in general requiring hotter, less dense plasma for lower Pedersen conductance.  The plot also confirms the results shown above, i.e.\ that typically jovian values for the source plasma population parameters and the Pedersen conductance lead to observed UCD radio luminosities.  \\

We now consider the effect of the form of the angular velocity profile, and show in Figure~\ref{fig:params} the radio luminosities obtained when the parameters in equation~\ref{eq:omega} are varied individually.  Specifically, from top to bottom the parameters varied are (a) the low value to which the angular velocity transitions from rigid corotation $(\omega / \Omega_p)_L$, (b) the latitudinal half-width of the transition region $\Delta\theta_i$, and (c) the centre of the transition region $\theta_{ic}$.  As a guide, the horizontal dotted lines highlight $1\;\mathrm{MW\;Hz^{-1}}$, which delimits the parameter ranges reasonably consistent with observations. Thus, an angular velocity profile which transitions to $\lesssim70$\% of rigid corotation across a region of half-width $\lesssim2.5^\circ$ at a co-latitude $\gtrsim9^\circ$ will produce radio powers consistent with observations.  These are not particularly tight constraints, indicating that the model does not require fine tuning to produce the required emissions. Future modelling of the radio beams, along the lines of previous work that has concentrated on localised active regions \citep{kuznetsov:2012aa}, from angular velocity transition regions of different geometries would be useful to constrain the angular velocity profiles further.\\

\begin{table*}
\begin{center}
\caption{Properties of the UCDs considered in this paper.\label{tab:ucds}}
\begin{tabular}{lccc}
\tableline\tableline
Property & TVLM 513 & LSR J1835 & 2M J0036 \\
\tableline
Radius $R_{UCD}$ / $\mathrm{R_J}$ \tablenotemark{a}& 1.005 & 1.142 & 0.927 \\
Rotation period $\tau$ / h \tablenotemark{a}& 1.96 & 2.84 & 3.08 \\
Polar ionospheric field strength $B_i$ / T  \tablenotemark{b}& 0.3 & 0.3 & 0.17 \\
Distance $d$ / pc  \tablenotemark{a} & 8.8 & 5.7 & 10.6 \\
Quiescent spectral luminosity $L_{r\:q}$ / $\mathrm{MW\;Hz^{-1}}$  \tablenotemark{c}& 5.77 & 2.77 & 1.23 \\
Mean peak spectral flux density $F_r$ / mJy & 4.2\tablenotemark{d}& 2.4\tablenotemark{e} & 0.5\tablenotemark{f}\\
Pulse duty cycle $\Delta \tau / \tau$ & 0.05\tablenotemark{g}& 0.1\tablenotemark{e} & 0.3\tablenotemark{f}\\
Estimated peak spectral luminosity $L_{r}$ / $\mathrm{MW\;Hz^{-1}}$  \tablenotemark{h} & 2.6 & 0.8 & 1.3 \\
Pedersen conductance $\Sigma_P^*$ / mho \tablenotemark{i} & 0.5 & 0.3 & 0.5 \\

\tableline
\end{tabular}
\tablenotetext{a}{From Table~1 of \cite{hallinan:2008aa}}
\tablenotetext{b}{From \cite{hallinan:2008aa}}
\tablenotetext{c}{From Table~2 of \cite{hallinan:2008aa}}
\tablenotetext{d}{From Figure~1 of \cite{hallinan:2007aa}}
\tablenotetext{e}{From Figure~1 of \cite{hallinan:2008aa}}
\tablenotetext{f}{From Figure~2 of \cite{hallinan:2008aa}}
\tablenotetext{g}{From Figure~3 of \cite{hallinan:2007aa}}
\tablenotetext{h}{From equation~\ref{eq:lr}}
\tablenotetext{i}{Employed in Figure~\ref{fig:cases}} 
\end{center}
\end{table*}

\subsection{Comparison with observations}

We now compare the model spectral luminosities with the confirmed sources of CMI-induced radio emissions discussed by \cite{hallinan:2008aa}, i.e.\ TVLM 513-46546, LSR J1835+3259, and 2MASS J00361617+1821104 (hereafter TVLM 513, LSR J1835 and 2M J0036), which have relevant properties listed in Table~\ref{tab:ucds}. We estimate the spectral luminosities of the CMI-induced radio emission from the peak amplitudes and widths plotted by \cite{hallinan:2007aa, hallinan:2008aa}.  As discussed above, CMI-induced radio emission is strongly beamed, into 1.6~sr for the case of Jupiter's HOM and DAM emissions, resulting in a duty cycle of $\sim$0.14 for these emissions.  We estimate the beam width $\sigma$ for these UCDs by simply scaling to the jovian emissions via

\begin{equation}
	\sigma = \left(\frac{\Delta\tau / \tau}{0.14}\right) 1.6\;\;,
	\label{eq:beam}
\end{equation}

\noindent where the pulse duty cycle $\Delta\tau / \tau$ is given in Table~\ref{tab:ucds}.  This, however, does assume that the angular velocity profile, and thus that of the field-aligned currents, are the same for each dwarf, which may not be the case.  The spectral luminosity then follows from the peak flux density $F$ via

\begin{equation}
	L_r = F\sigma d^2\;\;,
	\label{eq:lr}
\end{equation}

\noindent where $d$ is the distance to the object from the Earth.  This is shown by the crosses in Figure~\ref{fig:cases}.  In comparison, the modelled spectral luminosity, shown by the pluses, is computed using the radii, rotation periods, polar ionospheric field strengths and Pedersen conductances as listed in Table~\ref{tab:ucds}.  The Pedersen conductances employed for TVLM 513 and 2M J0036 are 0.5~mho, as is also used in Figure~\ref{fig:profs}, but it was found that 0.3~mho yields better agreement for LSR J1835.  It is also worth noting that employing 0.8~mho for all three dwarfs yields spectral luminosities consistent to within $\sim7$\% of the quiescent values (computed assuming isotropic emission) given by \cite{hallinan:2008aa}.  \\

We also note the potential CMI-induced emissions from the L dwarf binary 2MASSW J0746425+ 200032 \citep{berger:2009aa} and the T6.5 brown dwarf 2MASS J10475385+2124234 \citep{route:2012aa}. For simplicity we do not consider the  dwarf binary since the behaviour of such a system could be significantly more complex than is represented by our simple axisymmetric model.  Regarding the latter dwarf, two highly polarised bursts were observed at 4.75~GHz (implying a field strength of $\sim$0.17~T), with spectral flux densities of 2.7~$\pm$~0.2~mJy (where 1 Jansky = $10^{-26}\;\mathrm{W\;m^{-2}\;Hz^{-1}}$) and 1.3~$\pm$~0.2 mJy, respectively.  Assuming the radio emission is beamed into 1.6~sr and noting the object's 10.3~pc distance, these flux densities imply a mean spectral luminosity of $(3.3 \pm 0.7)~\mathrm{MW\;Hz^{-1}}$.  This observation is not included in Figure 2 since no rotation period is known, although it is likely to be at least 2~h since only one burst was observed in each $\sim$2~h observing interval.  Using the parameter values employed in this paper, our model indeed requires rotation periods of 2.1-2.8~h to produce radio powers consistent with the above luminosities, such that further observations of this object are required to test this scenario. \\

\section{Summary and Discussion}

In this paper we have developed an axisymmetric model of the M-I coupling currents arising from departure from rigid corotation in the magnetospheres of UCDs, based on that used previously to study Jupiter's auroral oval.  This simple model generates radio emissions whose power, bandwidth, modulation period and duty cycle are consistent with those observed.  The implication is that these UCDs are coupling with their space environments via magnetic fields, and that this coupling reduces the angular velocity of the high latitude ionopsheric regions, thus generating intense auroral emissions including radio emissions. \\

In this study the angular velocity profile is used as the input to the model, whereas in reality the form of the profile will be dictated by the physical processes governing the magnetospheric flows.  The magnetospheres of solar system planets are driven by two sources of momentum, i.e.\ the solar wind and planetary rotation.  Which of these is dominant depends on the individual circumstances, with Earth's magnetosphere being dominated by the solar wind, while Jupiter's and Saturn's are rotationally dominated. Previous work \citep{kuznetsov:2012aa} has assumed that the radio from UCDs is emitted from field lines mapping to $\rho_e = 2R_{UCD}$, where $\rho_e$ represents the equatorial radial distance from the magnetic axis, on the basis that at this distance the gravitational energy density equals the corotational energy density, such that it is hypothesised that beyond this distance any initially corotating plasma becomes centrifugally unstable and flows approximately radially away from the object \citep{ravi:2011aa}.  It is worth noting, however, that at Jupiter this distance is also $\sim$2~$\mathrm{R_J}$, and yet the dominant plasma flow is azimuthal out to the magnetopause at $\sim$40--100~$\mathrm{R_J}$ and no auroral emission is generated on field lines mapping to $\rho_e\simeq2\;\mathrm{R_J}$.  \\

A more complete equation of motion is

\begin{equation}
	\rho \left(\frac{d\mathbf{v}}{dt} - \mathbf{g}\right) +\nabla p = 
		\mathbf{j} \times \mathbf{B}\;\;,
	\label{eq:motn}
\end{equation}

\noindent where $\rho$ is the plasma mass density, $d\mathbf{v}/dt$ is the acceleration of the plasma bulk flow with respect to inertial space, $\mathbf{g}$ is the acceleration due to gravity, $p$ is the plasma pressure, $\mathbf{j}$ is the current density and $\mathbf{B}$ the magnetic field, and we note that in the corotating frame the $d\mathbf{v}/dt$ term comprises the convective and local accelerations, centrifugal, Coriolis and differential rotation effects \citep{vasyliunas83}.  The net result of the above force balance is that Jupiter's magnetosphere is radially distended into a magnetodisc structure, which doubles its size from that which might be expected from the planetary dipole alone \citep{caudal86,nichols11b}.  Plasma in Jupiter's magnetosphere is, however, centrifugally unstable and diffuses radially outward via flux tube interchange motions \citep{bespalov06} before being lost down the tail via the pinching off of plasmoids at around $\sim$100~$\mathrm{R_J}$ \citep{vogt10a}.  If there were no ionospheric torque on the magnetospheric plasma its angular velocity would decrease with  $1/\rho^2$, where $\rho$ is the distance from the spin axis, but the finite ionospheric Pedersen conductance allows a large-scale M-I coupling current system such as that discussed in this paper to flow, the $\mathbf{j} \times \mathbf{B}$ force of which balances the neutral drag in the ionosphere and maintains partial corotation throughout the closed magnetosphere.  The equation of motion which describes the angular velocity of the plasma under steady outflow from an internal source was derived originally by \cite{hill79}, and is given by


\begin{eqnarray}
	\lefteqn{\frac{\rho_e}{2}\frac{d}{d\rho_e}\left(\frac{\omega}{\Omega_{UCD}}\right)+\left(\frac{\omega}{\Omega_{UCD}}\right)=}\nonumber \\ &\frac{4\pi \Sigma_P^*F_e|B_{ze}|}{\dot{M}}\left(1-\frac{\omega}{\Omega_{UCD}}\right)\;\;,
	\label{eq:hp}
\end{eqnarray}

\noindent  where $F_e$ is the equatorial value of the the poloidal flux function $F$, related to the magnetic field $\mathbf{B}$ by $\mathbf{B}=(1/\rho)\nabla F \times \hat{\varphi}$, $|B_{ze}|$ is the magnitude of the north-south magnetic field threading the equatorial plane, and $\dot{M}$ is the plasma mass outflow rate.  Analytic solutions to equation~\ref{eq:hp} can be obtained, depending on the form of the magnetic field, along with characteristic distances over which the plasma breaks from rigid corotation \citep{nichols03}, called the `Hill distance' (not to be confused with the radius of the Hill sphere within which a body's gravitational force dominates).  In the absence of knowledge of the magnetodisc structure of these UCDs, taking the dipole case represents a lowest order approximation, for which the Hill distance is given by 

\begin{equation}
	\left(\frac{\rho_{H}}{R_{UCD}}\right) = \left(\frac{2\pi\Sigma_P^*B_{eq}^2 R_{UCD}^2}{\dot{M}}\right)^{1/4}\;\;,
	\label{eq:rhoh}
\end{equation}

\noindent where $B_{eq}$ is the equatorial field at the surface of the UCD.  Employing $\Sigma_P^*=0.5$~mho and the canonical value of $\dot{M} = 1000\;\mathrm{kg\;s^{-1}}$ for the Io source rate, we then have $\rho_H \simeq775\;\mathrm{R_{UCD}}$.  \\

In comparison, the size of the magnetosphere $R_{mp}$ is estimated by considering pressure balance between the magnetic field pressure of the compressed planetary dipole just inside the magnetopause and the magnetic, thermal, and dynamic pressures of the interstellar medium on the outside, i.e.\ 

\begin{equation}
	\left(\frac{\ensuremath{R_{mp}}}{R_{UCD}}\right)=\left(\frac{k_m^2B_{eq}^2}{2\mu_\circ (p_{th} + p_{dyn} + p_{B})}\right)^{1/6}\;\;,
	\label{eq:rmp}
\end{equation}

\noindent where $k_m$ is the factor by which the magnetopause field is enhanced by magnetopause currents, given by $\ensuremath{\sim}2.44$ for a frontal boundary of realistic shape, i.e.\ lying between the values of 2 and 3 appropriate to planar and spherical boundaries, respectively  \citep{mead64a,alexeev05a}.  Taking plasma number density, temperature and magnetic field strength values for both the local interstellar cloud and the local bubble, along with a representative velocity for the UCD of 50~$\mathrm{km\;s^{-1}}$ with respect to the local medium \citep{vanhamaki:2011aa}, yields values of $R_{mp} \simeq 713$ and $820~\mathrm{R_{UCD}}$ for the local interstellar cloud and local bubble, respectively.  Thus, on this basis corotation breakdown may only be marginal inside the closed magnetosphere, although a more detailed analysis of the force balance should be undertaken in the future to determine whether this would be the case for a realistic current sheet field structure.  \\

It is also worth mentioning that while at least TVLM 513 may possibly have a companion located at $20-50\;\mathrm{R_{UCD}}$ \citep{forbrich:2009aa}, evidence for satellite-induced radio emission has not been found \citep{kuznetsov:2012aa}, but in this regard we note that at Jupiter the brightest and most significant auroral form resulting from the Io plasma source is the main oval, whose apparent modulation is at the planetary period, not Io's orbital period.  If near-rigid corotation is then maintained throughout the closed magnetosphere, an interesting point is that the linear velocities of the plasma near the magnetopause may reach a significant fraction of the speed of light, i.e.\ $\sim$$0.17c$ for rigid corotation with a period of 2~h at $800\;\mathrm{R_{UCD}}$.  An order of magnitude approximation of the limiting distance at which plasma content can be maintained in corotation by inward magnetic stress is where the Alfv\'en speed equals the corotation speed.  The former is, of course, unknown since we do not have knowledge of the plasma density, but at high latitudes in Jupiter's magnetosphere it can approach the speed of light.  If the limiting distance does lie significantly within the magnetopause then a super-Alfv\'enic radially-outward planetary wind may indeed form, at which point the distinction between open and closed field lines could  become rather blurred \citep{kennel77}, although it is worth noting that at Jupiter this distance is also within the magnetosphere and no swift radial outflow occurs at this location.  \\

The amount of flux `truly' open to the interstellar medium will depend on the reconnection rates on the leading and trailing sides of the flow with respect to the surrounding medium.  Theories of magnetic reconnection have not yet reached the point where ab initio calculations of the rate of flux transport can be made, however it is known that reconnection occurs where a significant magnetic shear is present.  Without in situ measurements we do not of course know what the interstellar magnetic field is at the magnetopause, but it is likely to be piled up against the boundary, since the $\sim$50 $\mathrm{km\;s^{-1}}$ proper motions typical of low mass stars are similar to the velocities observed in the shocked solar wind plasma in Jupiter's magnetosheath just upstream of the magnetopause \citep[e.g. ][]{siscoe:1980aa}. At planets in the solar system the open-closed field line boundaries typically lie between $\sim$10--15$^\circ$ co-latitude.  The angular velocity of the open field line region \citep{isbell84} is given by

\begin{equation}
	\left(\frac{\omega}{\Omega_{UCD}}\right)_{Open} = 
		\frac{\mu_0 \Sigma_P^*v_{LIM}}{1 + \mu_0\Sigma_P^* v_{LIM}}\;\;,
	\label{eq:isbell}
\end{equation}

\noindent where $v_{LIM}$ is the velocity of the UCD through the local interstellar medium, which for the above parameter values yields $(\omega/\Omega_{UCD})_{Open}$ = 0.05, and may be much lower in the polar region if there is little aurorally-generated conductivity.  Therefore, an open-closed field line boundary is a possible cause of the angular velocity gradient discussed in this paper.  We note that \cite{schrijver:2009aa} considered the  interaction of the magnetosphere with the interstellar medium, and ruled it out, by showing that the kinetic energy flux of the impinging medium is much less than typical X-ray and H-$\alpha$ luminosities.  However, this does not consider the rotation of the body, which is a crucial property of these UCDs as discussed above.  Here we postulate that the interaction may instead be such that magnetic reconnection occurs on the front side of the magnetosphere and creates an essentially stagnant region of open field lines, whose angular velocities are given by equation~\ref{eq:isbell}, and the transition from corotating closed flux to sub-rotating open field lines drives field-aligned currents.  This is the current system described for Saturn's auroras by e.g. \cite{cowley03d} and \cite{cowley04a} and possibly a component of Jupiter's auroras poleward of the main oval \citep{cowley03f,cowley05a,cowley07}. Finally, it is worth noting that the auroral and radio emissions in our solar system caused by the above-discussed processes are all highly variable in terms of power, morphology and, in Saturn's case, modulation period, due to a number of reasons including varying interplanetary medium parameters and internal plasma source rates, such that it may be expected that the radio emissions from UCDs are similarly variable.


\acknowledgments

JDN was supported by an STFC Advanced Fellowship. SWHC was supported by STFC grant ST/H002480/1.  JDN wishes to thank G. Hallinan for helpful discussions regarding the radio observations of UCDs.


\clearpage

\clearpage


\end{document}